\documentclass[twocolumn]{aastex61}

\shorttitle{Oscillations Accompanying a Negative Flare II}
\shortauthors{N.~Kobanov and A.~Chelpanov}
\begin{document}
\title{Oscillations Accompanying a He\,\textsc{I} 10830\,\AA\ Negative Flare in a Solar Facula II. Response of the Transition Region and Corona}
\correspondingauthor{A.A.~\surname{Chelpanov}}
\email{chelpanov@iszf.irk.ru}
\author{Nikolai~\surname{Kobanov}}
\affil{Institute of Solar-Terrestrial Physics
                     of Siberian Branch of Russian Academy of Sciences, Irkutsk, Russia}
\author{Andrei~\surname{Chelpanov}}
\affiliation{Institute of Solar-Terrestrial Physics
                     of Siberian Branch of Russian Academy of Sciences, Irkutsk, Russia}
\begin{abstract}

We studied oscillations related to flare SOL2012-09-21T02:19 in the transition region and chromosphere based on \textit{Reuven Ramaty High Energy Solar Spectroscopic Imager} (RHESSI) and \textit{Solar Dynamics Observatory} (SDO) data, as well as data from the ground-based \textit{Horizontal Solar Telescope}. We found that 2-minute oscillations triggered by the flare first appeared in the RHESSI channels then subsequently showed up in the SDO 171\,\AA\ and 304\,\AA\ channels and after that in the  chromospheric He\,\textsc{i} 10830\,\AA\ line. The delay of the chromospheric signal compared to the RHESSI signal is 7 minutes, which indicates that the wave perturbation propagated from the corona to the chromosphere. We used the sharp increase in 3- and 5-minute oscillations during the flare in the lower atmosphere to trace the propagation of the oscillation trains to the transition region and corona. The results show that the 171\,\AA\ channel signals lagged behind the photospheric and chromospheric signals by 200\,s on average. We suggest that we observed slow magnetoacoustic waves both in the case of 2-minute oscillations propagating downwards from the corona and in the case of 3- and 5-minute oscillations leaking to the corona from beneath.
\end{abstract}

\section{Introduction} \label{sec:intro}

In the recent decades, quasi-periodic pulsations (QPPs) in solar flares became the subject for many studies of different ranges in the electromagnetic spectrum \citep{Aschwanden87, Grechnev2003, Kislyakov06,Sych2009, NakariakovMelnikov09, Kupriyanova10, Nakariakov2016SSR, Inglis2016, Kupriyanova2019}. \citet{Pugh17,Dominique} showed the difficulties in revealing QPPs and proposed algorithms to improve their identification in flare time series. \citet{McLaughlin} discussed the problems of QPP modelling.

In the preceding article \citep{Article1}, hereafter Article\,I, we analysed the influence of the small-scale flare on the oscillations in the lower atmosphere of a facula. We showed that small flare SOL2012-09-21T02:19 that we identified as a negative flare due to a sharp increase in the He\,\textsc{i} 10830\,\AA\ line absorption caused a deep modulation of the oscillation regime from the photosphere to the upper chromosphere. This impact was registered in the magnetic field strength, Doppler velocity, intensity, and line profile half-widths. This influence notably differed (up to the modulation sign) in different parts of the facula region.

The question of how the flare influenced the oscillation characteristics in the upper solar atmosphere was left out of the subject matter in Article\,I.

The aim of this article is to study the transition region and coronal oscillations related to the flare as well as to try to detect wave disturbances propagating both from the corona to the chromosphere and in the reverse direction. Note that our measurement capabilities drastically decrease compared to the lower atmospheric layers. In the transition region and corona, we can detect the oscillation and wave processes mostly based on the intensity variations in the \textit{Atmospheric Imaging Assembly} (AIA) signals.

\section{Instrument and methods}

To study flare SOL2012-09-21T02:19, we used the data from the \textit{Horizontal Solar Telescope} \citep[HST,][]{Kobanov01, Kobanov13} at the Sayan Solar Observatory, the \textit{Solar Dynamics Observatory} \citep[SDO,][]{Pesnell12}, and the \textit{Reuven Ramaty High Energy Solar Spectroscopic Imager} \citep[RHESSI,][]{rhessi1}.

The spectral observations of the Si\,\textsc{i} 10827\,\AA\ and He\,\textsc{i} 10830\,\AA\ lines were carried out with the HST. The telescope is built at the height of 2000\,m above sea level. It is equipped with a 900-mm main mirror. The focal length of the mirror is 20\,m.  A photoelectric guiding system stably positions an observed object with an accuracy of under 1$''$. The 6-meter spectrograph allows to measure line-of-sight (LOS) velocity, intensities, line-widths, and magnetic fields at a wide variety of spectral lines in the visible and infrared ranges \citep{Kobanov13}. The spectrograph is equipped with two mirror systems, which allows to record the spectral line characteristics in two different parts of the spectrum. The slit of the spectrograph comprises 65$''$ on the solar surface. The spectral resolution is 10\,m\AA\ per pixel on average. The time cadence of the series is 1.5\,seconds. The description of the telescope equipment and measurement methods can be found in articles \citet{Kobanov90,Kobanov01}.

The \textit{Atmospheric Imaging Assembly} \citep[AIA,][]{sdoaia} on board SDO continuously provides full-disk images in EUV channels with a 12-second cadence and UV channels with a 24-second cadence. The pixel size of the AIA images corresponds to 0.6$''$. We also used the X-ray flux recorded by the RHESSI.

The HST LOS velocity signals were restored using the lambdameter technique \citep{Rayrole, NegFlare2018}. The accuracy of the LOS velocity signal is 20\,m\,s$^{-1}$. The line intensities were normalized to the adjacent continuum intensities in order to compensate for the atmospheric weather condition changes. We applied the Fast Fourier Transform Interactive Data Language (IDL) algorithm to analyse the oscillation spectra of the signals. In order to avoid edge artifacts, we used the bell-shaped window function to 0.1 series length at each end of the signals. The signals were filtered in the given frequency ranges using the Morlet wavelet of the sixth order. To corroborate the presence of the 2-minute variations in the signals, we used the Empirical Mode Decomposition (EMD) technique \citep{Kolotkov16}.

\section{Results}

We analysed oscillations in the transition region and corona directly above the negative flare location. To do this, we use the SDO/AIA 304\,\AA, 171\,\AA\ channels along with the RHESSI 3--6\,keV and 6--12 keV channels---the two channels, whose intensity increased during the flare. The AIA channel signals were averaged over a 3$''$ by 3$''$ area in order to minimize possible projection effects.

\subsection{Propagation of the Flare-Related QPPs from the Corona to the Chromosphere}

\begin{figure}
\centerline{
\includegraphics[width=10cm]{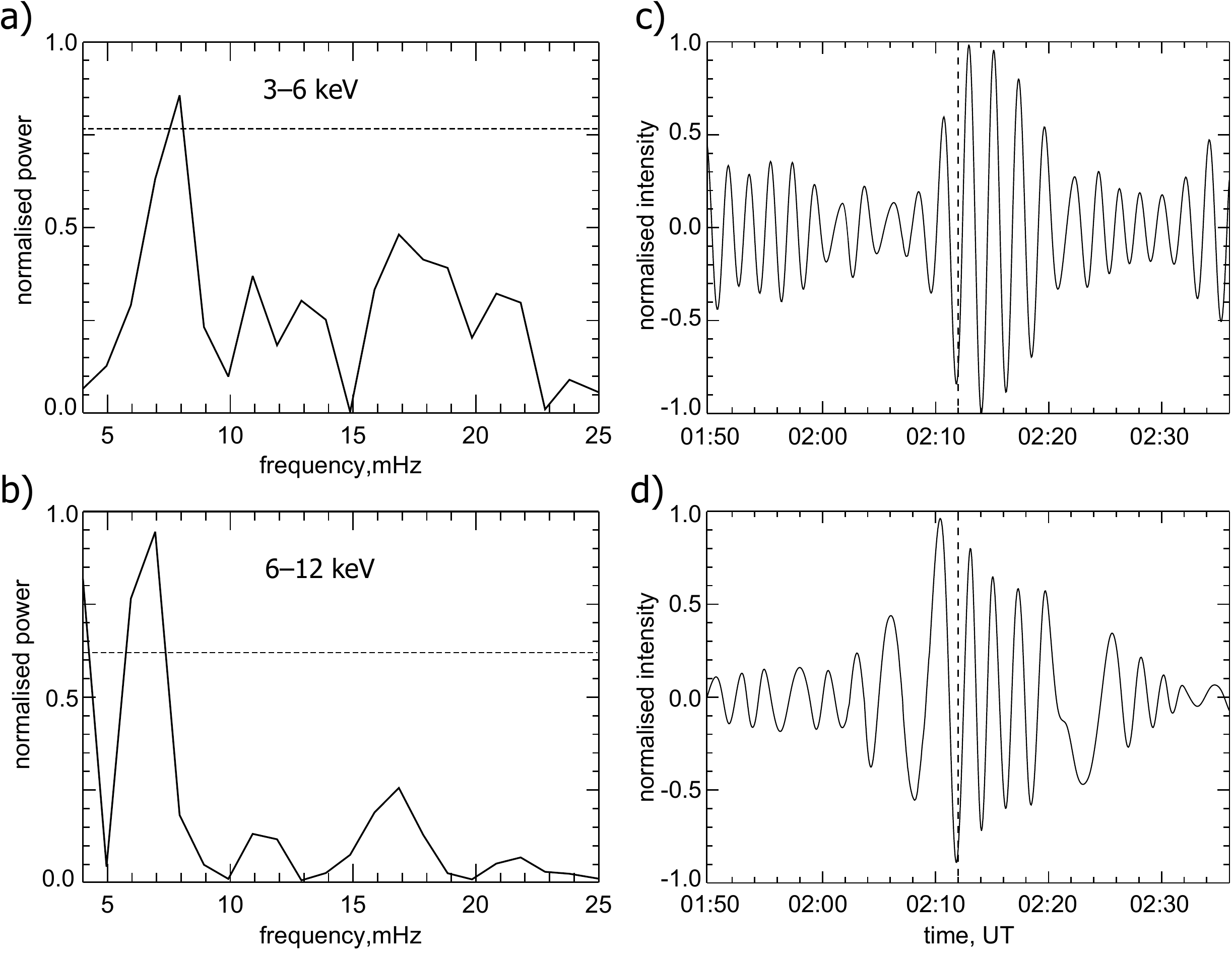}
}
\caption{X-ray flux oscillations: power spectra in the a) 3--6 and b) 6--12\,keV RHESSI channels during flare SOL2012-09-21T02:19; c) the 3--6\,keV channel signal filtered in the 8$\pm$0.7\,mHz range; d) 3--6\,keV channel signal EMD mode with a period close to 2 minutes. The oscillations hereinafter are normalized to 1. The dashed line in panels a and b shows the 90\% significance level.}
\label{fig:RhessiSpec}
\end{figure}

QPPs with a period close to two minutes were registered simultaneously in the 3--6\,keV and 6--12\,keV RHESSI channels, which supports the solar origin of these pulsations (Figure~\ref{fig:RhessiSpec}). The two-minute period also increases during the flare in the wavelet-filtred series and in the EMD mode, whose period is close to 2 minutes.

Note that oscillations with the 3- and 5-minute periods are more typical of the underlying photosphere and chromosphere, and the 2-minute period is observed rarely, if at all, in faculae. In such circumstances, the presence of the oscillations with a 2-minute period in the lower layers can confidently indicate that the oscillations triggered by the flare propagate from the corona to the chromosphere.

\begin{figure}
\centerline{
\includegraphics[width=6.5cm]{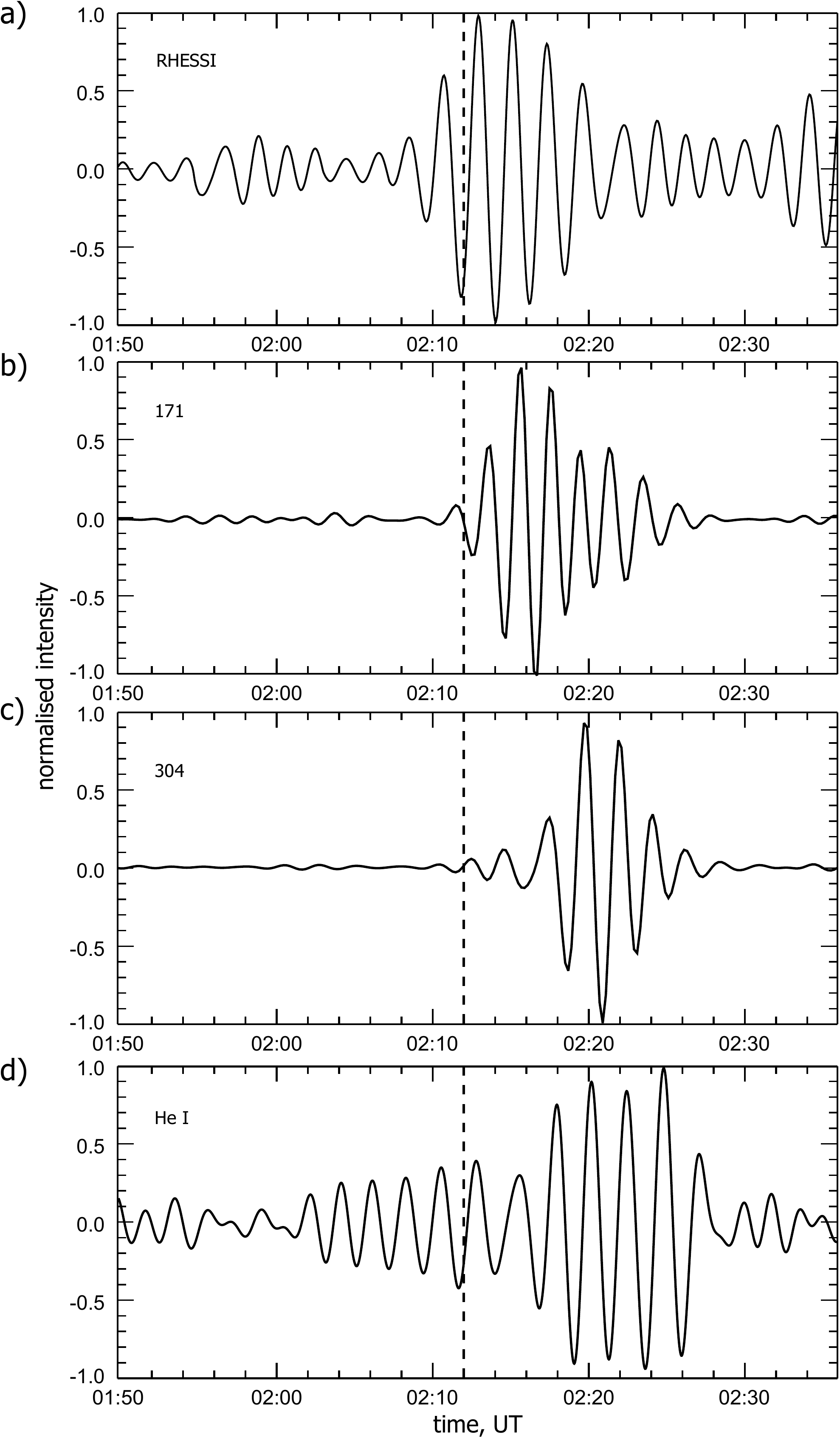}
}
\caption{Signals filtered in the 8.3$\pm$0.5\,mHz frequency range (2-minute oscillations) shown in the downward order: a) RHESSI 3--6\,keV channel; b) AIA\,171\,\AA\ channel; c) AIA\,304\,\AA\ channel; d) He\,\textsc{i} 10830\,\AA\ line. The vertical dashed line hereinafter shows the start of the flare in the X-ray data.}
\label{fig:2min}
\end{figure}

The AIA 171\,\AA\ and 304\,\AA\ channel signals, as well as the He\,\textsc{i} 10830\,\AA\ spectral data support the fact that 2-minute oscillations propagate downwards from the corona to the lower layers. Figure \ref{fig:2min} clearly shows that the signals of the lower levels lag behind those of the upper ones. The centers of the wave trains gradually shifts from 02:15 in the RHESSI signal to 02:22 in the He\,\textsc{i} 10830\,\AA\ line signal. The wave train in the 171\,\AA\ channel follows 180$\pm$20\,s after the wave train in the RHESSI data, and 180$\pm$25\,s after that it appears in the 304\,\AA\ channel. The center of the main wave train in the He\,\textsc{i} 10830\,\AA\ line lags 60$\pm$15\,s behind that of the 304\,\AA\ channel.

The signals that we measure at different levels are averaged over 3$''$ by 3$''$ areas, whose centers are aligned perpendicular to the solar surface. In order to examine whether the signal diverted from the vertical propagation at the 304\,\AA\ level, we measured the signals in four similar areas displaced in different direction by 3.6$''$ from the original location. The amplitude of the signals in these areas was 4--5 times lower. In our opinion, this indicates that we measured the signals in the main propagation channel of the studied waves. Thus, at the transition region---chromosphere---photosphere height range this channel is close to vertical.

In addition, earlier in the analysis  of this flare, \citet{Article1} showed that the velocity of the perturbation propagation was 70\,km\,s$^{-1}$ in the image plane at the 171\AA\ level.

\subsection{Penetration of the Photospheric 5-minute and Chromospheric 3-minute Oscillations to the Corona}

As Article\,I showed, the amplitudes of 3- and 5-minute chromospheric oscillations drastically increased during the flare. The signals look as increased amplitude oscillation trains, which appeared simultaneously with the flare. Such a train structure of the signals serves as a useful means to study propagating oscillations in the solar atmosphere. Note that in the absence of flare perturbations, intrinsic oscillations of these periods, though with lower amplitudes, may occur continuously in both active and quiet regions. In this case, however, estimates of the phase relations in the propagating signals show more ambiguity. While the direct phase lag measurements can be straightforwardly carried out at the lower levels, this poses difficulties in the transition region and corona \citep{ Prasad15, Zhao16}. \citet{Prasad15} demonstrated a signal amplitude modulation, which helped them trace the propagation of \textit{p}-modes from the photosphere to the corona in a form of slow magnetoacoustic waves.

\begin{figure}
\centerline{
\includegraphics[width=6cm]{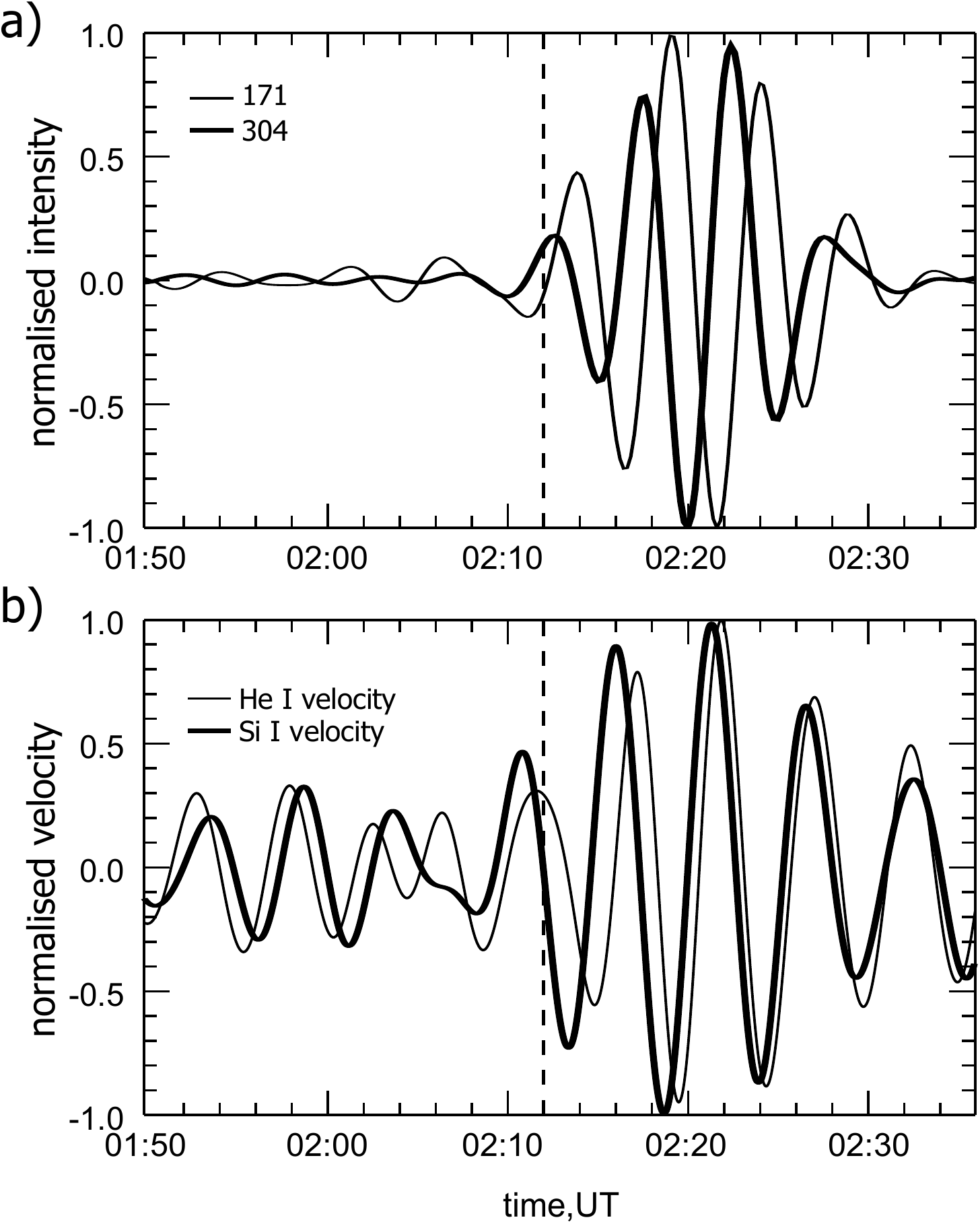}
}
\caption{Signals filtered in the 3.3$\pm$0.5\,mHz frequency range (5-minute oscillations): a) AIA\,304\,\AA\ and 171\,\AA\ channel intenssities; b) Si\,\textsc{i} 10827\,\AA\ and He\,\textsc{i} 10830\,\AA\ line LOS velocity.}
\label{fig:5min}
\end{figure}

Figure~\ref{fig:5min}b shows 5-minute oscillations in the LOS velocity signals of the Si\,\textsc{i} 10827\,\AA\ (temperature minimum) and He\,\textsc{i} 10830\,\AA\ (upper chromosphere) lines. The chromospheric oscillation train follows the photospheric one 60\,s later on average. As seen in Figure~\ref{fig:5min}b, the phase difference notably decreases by the end of the oscillation train. This might be due to the change in the oscillation regime from the running acoustic waves to standing waves in the upper photosphere. Figure~2b in Article\,I supports this explanation: the phase shift between the intensity and LOS velocity oscillations of the Si\,\textsc{i} 10827\,\AA\ line changes from 180$^\circ$ to 90$^\circ$ by the end of the flare. The oscillation train of the AIA 304\,\AA\ channel (Figure~\ref{fig:5min}a) almost coincides with that of the He\,\textsc{i} line in panel b. In turn, 5-minute oscillations in the 171\,\AA\ channel lag behind those in the 304\,\AA\ by approximately 100\,s. These time relations indicate that 5-minute oscillations propagated from the facular photosphere through the chromosphere and transition region to the lower corona.

\begin{figure}
\centerline{
\includegraphics[width=6cm]{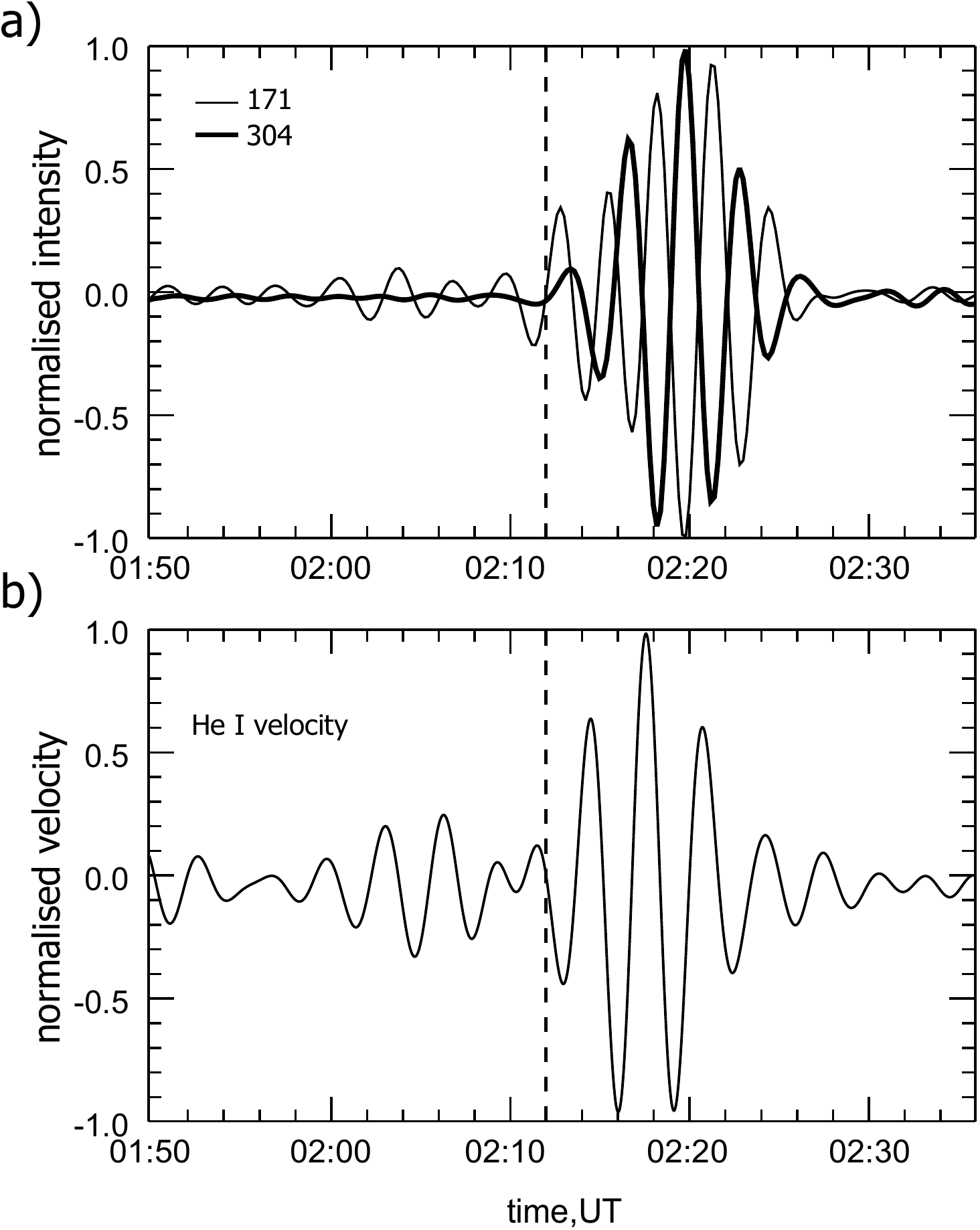}
}
\caption{Signals filtered in the 5.5$\pm$0.5\,mHz frequency range (3-minute oscillations): a) AIA\,304\,\AA\ and 171\,\AA\ channels; b) He\,\textsc{i} 10830\,\AA\ line LOS velocity.}
\label{fig:3min}
\end{figure}

The 3-minute oscillation trains in the 304\,\AA\ and 171\,\AA\ channels lag behind the chromospheric oscillations in the He\,\textsc{i} line by 120$\pm$20 and 200$\pm$28\,s, respectively (Figure~\ref{fig:3min}). As in the case of 5-minute oscillations, this clearly indicates that 3-minute oscillations propagate from the chromosphere to the corona.

In this article, we chose not to measure the propagation speeds of the observed wave disturbances in the corona for the following reasons: first, the height scale of the SDO/AIA channels is difficult to define in flaring regions; second, the real propagation trajectory of different oscillation modes differ and may notably diverge from the vertical direction. We, however, do it for the height interval between the temperature minimum (Si\,\textsc{i} 10827\,\AA) and upper chromosphere (He\,\textsc{i} 10830\,\AA). According to \citet{BardCarlsson08, Leenaarts12}, the difference in the formation heights of these two lines may reach 800--900\,km. Thus, based on the direct measurement of the signal phase delay, we can determine its propagation speed \citep{Kobanov13SP}. The time lag between the signals is 60\,s on average, which yields the propagation speed of the 5-minute oscillations to be 13--15\,km\,s$^{-1}$.

\section{Discussion}

QPPs with a 2-minute period are observed rarely compared to those with sub-minute periods. \citet{Kuznetsov16} concluded that such QPPs are closely related to eruptive events.

\citet{Liu11} in their analysis of the AIA 171\,\AA\ intensity variations in a C3-class flare revealed waves with the dominant period of 181\,s propagating along the coronal loop funnel. The measured phase velocity of these waves reached 2000\,km\,s$^{-1}$. The authors identified them as the fast magnetoacoustic waves. Earlier, \citet{Terekhov2002} found a period of 143.2\,s in a 8--20\,keV light curve of a flare. Before that, an idea that Alfv{\'e}n waves transport the flare energy from the corona to the lower solar atmosphere was proposed by \citet{Emslie82} and later developed by \citet{FletcherHudson08}. In this theory, the preference was given to torsional Alfv{\'e}n waves. This is hardly applicable in our case, since torsional Alfv{\'e}n waves are not accompanied by intensity oscillations in the He\,\textsc{i} 10830\,\AA\ chromospheric line shown in Figure~\ref{fig:2min}d. \citet{YuanShen2013} observed `quasi-periodic propagating fast magneto-acoustic wave trains (QPF)' 100\,Mm away from the C2-class flare site. QPF period was under 1\,min, and the propagation speed was 700--800\,km\,s$^{-1}$.

\citet{Ning} found a 2-minute period in a B8.1-class flare. The wave perturbation propagated along the flare loop at a speed of 130\,km\,s$^{-1}$, which is close to the local sound speed. \citet{Nistico2014} observed wave trains caused by a powerful flare. The measured period was close to 1\,min, and the propagation velocity exceeded 1000\,km\,s$^{-1}$.

\citet{ZhangZhang2015} showed that fast and slow magnetoacoustic waves can propagate by the same trajectories. They measured the period and velocity of the fast waves to be 2\,min and 900\,km\,s$^{-1}$, and those of the slow waves to be 3\,min and 100\,km\,s$^{-1}$. The authors, however, note that, as opposed to the fast wave trains, the slow waves most likely result from the chromosphere leakage rather than from the flare disturbance.

\citet{Brosius18} found periods around 2 minutes in the LOS velocity and intensity variations in a flare ribbon during chromospheric evaporation. \citet{Milligan17} observed two main periods of 120 and 180 minutes in an X-class flare, and these 2-minute oscillations were registered both in the hard X-ray and the chromospheric EUV radiation. The authors state that this is due to the chromospheric response to accelerated particles. The delays between the X-ray oscillations and the oscillations in different AIA channels were not estimated in this article.

The 2-minute oscillations that we observed were most likely generated by the small flare. Their spatial and temporal concurrence indirectly supports this suggestion.

Having analysed the characteristics of these oscillations we believe that thay can be related to slow magnetoacoustic waves.

The problem of the oscillation propagation from the chromosphere to the corona and in the reverse direction has existed for a long time. In the last decades, with the advent of the space telescope data it has prompted more interest. Nevertheless, due to the complexity of the problem, the conclusions of the early studies contain notable discrepancies. For example, \citet{DeMoortel02} found that oscillations with a period of 321$\pm$74\,s are observed above faculae in the \textit{Transition Region and Coronal Explorer} (TRACE) 171\,\AA\ channel intensity. They suggested that this results from photospheric oscillations propagating through the transition region to the corona. With this, the authors found no evidence for the downward propagation. In the analysis of similar TRACE 171\,\AA\ data, \citet{DePontieu03} concluded that 5-minute oscillations in the upper transition region do not result from a direct leakage of the \textit{p}-mode from the photosphere through the chromosphere. The authors believe that these oscillations are connected non-linearly. \citet{Centeno06,Khomenko08} provided observational evidence that 5-minute oscillations propagate from the photosphere to the chromosphere in the vertical magnetic tubes of faculae. The authors state that the cut-off frequency may significantly decrease due to radiative losses. However, \citet{deWijn09} based on the high-resolution \textit{Hinode} data, found that 3-minute oscillations propagate upwards in vertical magnetic tubes of faculae, while propagating 5-minute oscillations were observed only at facular peripheries in inclined magnetic fields. \citet{Chelpanov15} showed that in the AIA 304\,\AA\ channel intensity, the main oscillation power is distributed in the 3--6\,mHz range above facular magnetic knots and in the 1.5--3\,mHz above facular peripheries. They found oscillations at a frequency of 5\,mHz in the magnetic field strength signals in magnetic knots.

We suggest that, given such large time lags between the signals and the other characteristics, we observed slow magnetoacoustic waves \citep{Zimovets09,NakariakovZimovets11}. This refers to both 2-minute oscillations propagating downwards from the corona and 5- with 3-minute oscillations leaking from the photosphere and chromosphere to the lower corona (the 171\,\AA\ channel).

\section{Conclusion}

In the analysis of flare SOL2012-09-21T02:19, we found QPPs with a 2-minute period in the 3--6 and 6--12\,keV RHESSI channels. Similar variations were found in the SDO/AIA EUV channels and in the spectral data of the ground-based telescope in the He\,\textsc{i} 10830\,\AA\ line. These variations have a form of an oscillation train with increased amplitudes that lasted for 3--4 periods. The centers of the oscillation trains gradually shifted from 02:15\,UT in the RHESSI channels to 02:22\,UT in the He\,\textsc{i} 10830\,\AA\ line. The authors suggest that such a delay unambiguously indicates the propagation of the flare-triggered wave disturbance with a period of 2 minutes from the corona to the chromosphere.

We measured the 3- and 5-minute oscillation delays in their propagation from the photosphere and chromosphere to the transition region and corona. The 5-minute oscillation signal of the He\,\textsc{i} 10830\,\AA\ line lags 60\,s on average behind the Si\,\textsc{i} 10827\,\AA\ line signal, while the AIA\,304\,\AA\ and 171\,\AA\ signals lag 100\,s and 200\,s, respectively. Three-minute oscillations in the 304\,\AA\ and 171\,\AA\ channels lag 120\,s and 200\,s behind the chromospheric signals.

We suggest that the combination of the oscillation characteristics applies to the slow magnetoacoustic wave mode in a greater degree than to other modes.

The results show that the oscillation train structure of the signals resulting from an abrupt increase of the oscillation amplitudes in the lower atmosphere caused by a small-scale flare \citep{Article1} may serve as a useful means in studies of the wave propagation processes in the transition region and corona.

\acknowledgments

The research was partially supported by Project No.\,II.16.3.2 of ISTP SB RAS. Spectral  data were recorded at the Angara Multiaccess Center facilities at ISTP SB RAS. We acknowledge the NASA/SDO and RHESSI science teams for providing the coronal observation data. We are grateful to the anonymous referee for the useful remarks and suggestions.

%\clearpage\newpage
\bibliographystyle{spr-mp-sola}

\tracingmacros=2
\bibliography{Kobanov19}

\begin{thebibliography}{46}
% BibTex style file: spr-mp-sola.bst (nameyear), 2014-01-28
\ifx\bisbn     \undefined \def\bisbn  #1{ISBN #1}\fi
\ifx\binits    \undefined \def\binits#1{#1}\fi
\ifx\bauthor   \undefined \def\bauthor#1{#1}\fi
\ifx\batitle   \undefined \def\batitle#1{#1}\fi
\ifx\bjtitle   \undefined \def\bjtitle#1{\textit{#1}}\fi
\ifx\bvolume   \undefined \def\bvolume#1{\textbf{#1}}\fi
\ifx\byear     \undefined \def\byear#1{#1}\fi
\ifx\bissue    \undefined \def\bissue#1{#1}\fi
\ifx\bfpage    \undefined \def\bfpage#1{#1}\fi
\ifx\blpage    \undefined \def\blpage #1{#1}\fi
\ifx\burl      \undefined \def\burl#1{\textsf{#1}}\fi
\ifx\href      \undefined \def\href#1#2{\textsf{#2}}\fi
\ifx\betal     \undefined \def\betal{\textit{et al.}}\fi
\ifx\bctitle   \undefined \def\bctitle#1{#1}\fi
\ifx\beditor   \undefined \def\beditor#1{#1}\fi
\ifx\bbtitle   \undefined \def\bbtitle#1{\textit{#1}}\fi
\ifx\bedition  \undefined \def\bedition#1{#1}\fi
\ifx\bseriesno \undefined \def\bseriesno#1{\textbf{#1}}\fi
\ifx\blocation \undefined \def\blocation#1{#1}\fi
\ifx\bsertitle \undefined \def\bsertitle#1{\textit{#1}}\fi
\ifx\bsnm      \undefined \def\bsnm#1{#1}\fi
\ifx\bsuffix   \undefined \def\bsuffix#1{#1}\fi
\ifx\bparticle \undefined \def\bparticle#1{#1}\fi
\ifx\barticle  \undefined \def\barticle#1{}\fi
\ifx\binstitute  \undefined \def\binstitute#1{#1}\fi
\ifx\bpublisher  \undefined \def\bpublisher#1{#1}\fi
\ifx\doiurl    \undefined
  \def\doiurl#1{\href{http://dx.doi.org/#1}{\textsf{DOI}}}\fi
\ifx\arxivurl  \undefined
  \def\arxivurl#1{\href{http://arxiv.org/abs/#1}{\textsf{arXiv}}}\fi
\ifx\adsurl    \undefined
  \def\adsurl#1{\href{http://adsabs.harvard.edu/abs/#1}{\textsf{ADS}}}\fi
\ifx\botherref \undefined \def\botherref#1{}\fi
\ifx\url       \undefined \def\url#1{\textsf{#1}}\fi
\ifx\bchapter  \undefined \def\bchapter#1{}\fi
\ifx\bbook     \undefined \def\bbook#1{}\fi
\ifx\bcomment  \undefined \def\bcomment#1{#1}\fi
\ifx\oauthor   \undefined \def\oauthor#1{#1}\fi
\ifx\citeauthoryear \undefined\def \citeauthoryear#1{#1}\fi
\def\endbibitem {}
\ifx\bconflocation  \undefined \def\bconflocation#1{#1} \fi

\bibitem[\protect\citeauthoryear{{Aschwanden}}{1987}]{Aschwanden87}
\begin{barticle}
\bauthor{\bsnm{{Aschwanden}}, \binits{M.J.}}:
\byear{1987},
\batitle{{Theory of radio pulsations in coronal loops}}.
\bjtitle{\solphys}
\bvolume{111},
\bfpage{113}\,--\,\blpage{136}.
\doiurl{10.1007/BF00145445}.
\adsurl{1987SoPh..111..113A}.
\end{barticle}
\endbibitem

\bibitem[\protect\citeauthoryear{{Bard} and {Carlsson}}{2008}]{BardCarlsson08}
\begin{barticle}
\bauthor{\bsnm{{Bard}}, \binits{S.}},
\bauthor{\bsnm{{Carlsson}}, \binits{M.}}:
\byear{2008},
\batitle{{Constructing Computationally Tractable Models of Si I for the 1082.7
  nm Transition}}.
\bjtitle{\apj}
\bvolume{682},
\bfpage{1376}\,--\,\blpage{1385}.
\doiurl{10.1086/589910}.
\adsurl{2008ApJ...682.1376B}.
\end{barticle}
\endbibitem

\bibitem[\protect\citeauthoryear{{Brosius} and {Inglis}}{2018}]{Brosius18}
\begin{barticle}
\bauthor{\bsnm{{Brosius}}, \binits{J.W.}},
\bauthor{\bsnm{{Inglis}}, \binits{A.R.}}:
\byear{2018},
\batitle{{Localized Quasi-periodic Fluctuations in C II, Si IV, and Fe XXI
  Emission during Chromospheric Evaporation in a Flare Ribbon Observed by IRIS
  on 2017 September 9}}.
\bjtitle{\apj}
\bvolume{867},
\bfpage{85}.
\doiurl{10.3847/1538-4357/aae5f5}.
\adsurl{2018ApJ...867...85B}.
\end{barticle}
\endbibitem

\bibitem[\protect\citeauthoryear{{Centeno}, {Collados}, and {Trujillo
  Bueno}}{2006}]{Centeno06}
\begin{bchapter}
\bauthor{\bsnm{{Centeno}}, \binits{R.}},
\bauthor{\bsnm{{Collados}}, \binits{M.}},
\bauthor{\bsnm{{Trujillo Bueno}}, \binits{J.}}:
\byear{2006},
\bctitle{{Oscillations and Wave Propagation in Different Solar Magnetic
  Features}}.
In: \beditor{\bsnm{{Casini}}, \binits{R.}},
\beditor{\bsnm{{Lites}}, \binits{B.W.}} (eds.)
\bbtitle{Solar Polarization 4},
\bsertitle{Astronomical Society of the Pacific Conference Series}
\bseriesno{358},
\bfpage{465}.
\adsurl{2006ASPC..358..465C}.
\end{bchapter}
\endbibitem

\bibitem[\protect\citeauthoryear{{Chelpanov} and {Kobanov}}{2018}]{Article1}
\begin{barticle}
\bauthor{\bsnm{{Chelpanov}}, \binits{A.A.}},
\bauthor{\bsnm{{Kobanov}}, \binits{N.I.}}:
\byear{2018},
\batitle{{Oscillations Accompanying a He i 10830 {\AA} Negative Flare in a
  Solar Facula}}.
\bjtitle{\solphys}
\bvolume{293},
\bfpage{157}.
\doiurl{10.1007/s11207-018-1378-2}.
\adsurl{2018SoPh..293..157C}.
\end{barticle}
\endbibitem

\bibitem[\protect\citeauthoryear{{Chelpanov}, {Kobanov}, and
  {Kolobov}}{2015}]{Chelpanov15}
\begin{barticle}
\bauthor{\bsnm{{Chelpanov}}, \binits{A.A.}},
\bauthor{\bsnm{{Kobanov}}, \binits{N.I.}},
\bauthor{\bsnm{{Kolobov}}, \binits{D.Y.}}:
\byear{2015},
\batitle{{Characteristics of oscillations in magnetic knots of solar faculae}}.
\bjtitle{Astronomy Reports}
\bvolume{59},
\bfpage{968}\,--\,\blpage{973}.
\doiurl{10.1134/S1063772915090036}.
\adsurl{2015ARep...59..968C}.
\end{barticle}
\endbibitem

\bibitem[\protect\citeauthoryear{{De Moortel}
  \textit{et~al.}}{2002}]{DeMoortel02}
\begin{barticle}
\bauthor{\bsnm{{De Moortel}}, \binits{I.}},
\bauthor{\bsnm{{Ireland}}, \binits{J.}},
\bauthor{\bsnm{{Hood}}, \binits{A.W.}},
\bauthor{\bsnm{{Walsh}}, \binits{R.W.}}:
\byear{2002},
\batitle{{The detection of 3 and 5 min period oscillations in coronal loops}}.
\bjtitle{\aap}
\bvolume{387},
\bfpage{L13}\,--\,\blpage{L16}.
\doiurl{10.1051/0004-6361:20020436}.
\adsurl{2002A\%26A...387L..13D}.
\end{barticle}
\endbibitem

\bibitem[\protect\citeauthoryear{{De Pontieu}, {Erd{\'e}lyi}, and {de
  Wijn}}{2003}]{DePontieu03}
\begin{barticle}
\bauthor{\bsnm{{De Pontieu}}, \binits{B.}},
\bauthor{\bsnm{{Erd{\'e}lyi}}, \binits{R.}},
\bauthor{\bsnm{{de Wijn}}, \binits{A.G.}}:
\byear{2003},
\batitle{{Intensity Oscillations in the Upper Transition Region above Active
  Region Plage}}.
\bjtitle{\apjl}
\bvolume{595},
\bfpage{L63}\,--\,\blpage{L66}.
\doiurl{10.1086/378843}.
\adsurl{2003ApJ...595L..63D}.
\end{barticle}
\endbibitem

\bibitem[\protect\citeauthoryear{{de Wijn}, {McIntosh}, and {De
  Pontieu}}{2009}]{deWijn09}
\begin{barticle}
\bauthor{\bsnm{{de Wijn}}, \binits{A.G.}},
\bauthor{\bsnm{{McIntosh}}, \binits{S.W.}},
\bauthor{\bsnm{{De Pontieu}}, \binits{B.}}:
\byear{2009},
\batitle{{On the Propagation of p-Modes Into the Solar Chromosphere}}.
\bjtitle{\apjl}
\bvolume{702},
\bfpage{L168}\,--\,\blpage{L171}.
\doiurl{10.1088/0004-637X/702/2/L168}.
\adsurl{2009ApJ...702L.168D}.
\end{barticle}
\endbibitem

\bibitem[\protect\citeauthoryear{{Dominique} \textit{et~al.}}{2018}]{Dominique}
\begin{barticle}
\bauthor{\bsnm{{Dominique}}, \binits{M.}},
\bauthor{\bsnm{{Zhukov}}, \binits{A.N.}},
\bauthor{\bsnm{{Dolla}}, \binits{L.}},
\bauthor{\bsnm{{Inglis}}, \binits{A.}},
\bauthor{\bsnm{{Lapenta}}, \binits{G.}}:
\byear{2018},
\batitle{{Detection of Quasi-Periodic Pulsations in Solar EUV Time Series}}.
\bjtitle{\solphys}
\bvolume{293},
\bfpage{61}.
\doiurl{10.1007/s11207-018-1281-x}.
\adsurl{2018SoPh..293...61D}.
\end{barticle}
\endbibitem

\bibitem[\protect\citeauthoryear{{Emslie} and {Sturrock}}{1982}]{Emslie82}
\begin{barticle}
\bauthor{\bsnm{{Emslie}}, \binits{A.G.}},
\bauthor{\bsnm{{Sturrock}}, \binits{P.A.}}:
\byear{1982},
\batitle{{Temperature minimum heating in solar flares by resistive dissipation
  of Alfven waves}}.
\bjtitle{\solphys}
\bvolume{80},
\bfpage{99}\,--\,\blpage{112}.
\doiurl{10.1007/BF00153426}.
\adsurl{1982SoPh...80...99E}.
\end{barticle}
\endbibitem

\bibitem[\protect\citeauthoryear{{Fletcher} and
  {Hudson}}{2008}]{FletcherHudson08}
\begin{barticle}
\bauthor{\bsnm{{Fletcher}}, \binits{L.}},
\bauthor{\bsnm{{Hudson}}, \binits{H.S.}}:
\byear{2008},
\batitle{{Impulsive Phase Flare Energy Transport by Large-Scale Alfv{\'e}n
  Waves and the Electron Acceleration Problem}}.
\bjtitle{\apj}
\bvolume{675},
\bfpage{1645}\,--\,\blpage{1655}.
\doiurl{10.1086/527044}.
\adsurl{2008ApJ...675.1645F}.
\end{barticle}
\endbibitem

\bibitem[\protect\citeauthoryear{{Grechnev}, {White}, and
  {Kundu}}{2003}]{Grechnev2003}
\begin{barticle}
\bauthor{\bsnm{{Grechnev}}, \binits{V.V.}},
\bauthor{\bsnm{{White}}, \binits{S.M.}},
\bauthor{\bsnm{{Kundu}}, \binits{M.R.}}:
\byear{2003},
\batitle{{Quasi-periodic Pulsations in a Solar Microwave Burst}}.
\bjtitle{\apj}
\bvolume{588},
\bfpage{1163}\,--\,\blpage{1175}.
\doiurl{10.1086/374315}.
\adsurl{2003ApJ...588.1163G}.
\end{barticle}
\endbibitem

\bibitem[\protect\citeauthoryear{{Inglis} \textit{et~al.}}{2016}]{Inglis2016}
\begin{barticle}
\bauthor{\bsnm{{Inglis}}, \binits{A.R.}},
\bauthor{\bsnm{{Ireland}}, \binits{J.}},
\bauthor{\bsnm{{Dennis}}, \binits{B.R.}},
\bauthor{\bsnm{{Hayes}}, \binits{L.}},
\bauthor{\bsnm{{Gallagher}}, \binits{P.}}:
\byear{2016},
\batitle{{A Large-scale Search for Evidence of Quasi-periodic Pulsations in
  Solar Flares}}.
\bjtitle{\apj}
\bvolume{833},
\bfpage{284}.
\doiurl{10.3847/1538-4357/833/2/284}.
\adsurl{2016ApJ...833..284I}.
\end{barticle}
\endbibitem

\bibitem[\protect\citeauthoryear{{Khomenko} \textit{et~al.}}{2008}]{Khomenko08}
\begin{barticle}
\bauthor{\bsnm{{Khomenko}}, \binits{E.}},
\bauthor{\bsnm{{Centeno}}, \binits{R.}},
\bauthor{\bsnm{{Collados}}, \binits{M.}},
\bauthor{\bsnm{{Trujillo Bueno}}, \binits{J.}}:
\byear{2008},
\batitle{{Channeling 5 Minute Photospheric Oscillations into the Solar Outer
  Atmosphere through Small-Scale Vertical Magnetic Flux Tubes}}.
\bjtitle{\apjl}
\bvolume{676},
\bfpage{L85}.
\doiurl{10.1086/587057}.
\adsurl{2008ApJ...676L..85K}.
\end{barticle}
\endbibitem

\bibitem[\protect\citeauthoryear{{Kislyakov}
  \textit{et~al.}}{2006}]{Kislyakov06}
\begin{barticle}
\bauthor{\bsnm{{Kislyakov}}, \binits{A.G.}},
\bauthor{\bsnm{{Zaitsev}}, \binits{V.V.}},
\bauthor{\bsnm{{Stepanov}}, \binits{A.V.}},
\bauthor{\bsnm{{Urpo}}, \binits{S.}}:
\byear{2006},
\batitle{{On the Possible Connection between Photospheric 5-Min Oscillation and
  Solar Flare Microwave Emission}}.
\bjtitle{\solphys}
\bvolume{233},
\bfpage{89}\,--\,\blpage{106}.
\doiurl{10.1007/s11207-006-2850-y}.
\adsurl{2006SoPh..233...89K}.
\end{barticle}
\endbibitem

\bibitem[\protect\citeauthoryear{{Kobanov}, {Chelpanov}, and
  {Pulyaev}}{2018}]{NegFlare2018}
\begin{barticle}
\bauthor{\bsnm{{Kobanov}}, \binits{N.}},
\bauthor{\bsnm{{Chelpanov}}, \binits{A.}},
\bauthor{\bsnm{{Pulyaev}}, \binits{V.}}:
\byear{2018},
\batitle{{Negative flare in the He I 10830 {\AA} line in facula}}.
\bjtitle{Journal of Atmospheric and Solar-Terrestrial Physics}
\bvolume{173},
\bfpage{50}\,--\,\blpage{56}.
\doiurl{10.1016/j.jastp.2018.04.007}.
\adsurl{2018JASTP.173...50K}.
\end{barticle}
\endbibitem

\bibitem[\protect\citeauthoryear{{Kobanov}}{1990}]{Kobanov90}
\begin{barticle}
\bauthor{\bsnm{{Kobanov}}, \binits{N.I.}}:
\byear{1990},
\batitle{{On spatial characteristics of five-minute oscillations in the sunspot
  umbra}}.
\bjtitle{\solphys}
\bvolume{125},
\bfpage{25}\,--\,\blpage{30}.
\doiurl{10.1007/BF00154775}.
\adsurl{1990SoPh..125...25K}.
\end{barticle}
\endbibitem

\bibitem[\protect\citeauthoryear{{Kobanov}}{2001}]{Kobanov01}
\begin{barticle}
\bauthor{\bsnm{{Kobanov}}, \binits{N.I.}}:
\byear{2001},
\batitle{{Measurements of the differential line-of-sight velocity and
  longitudinal magnetic field on the Sun with CCD photodetector: part I.
  Modulationless techniques.}}
\bjtitle{Instrum. Exper. Tech.}
\bvolume{4},
\bfpage{110}\,--\,\blpage{115}.
\adsurl{2001InExT...4..110K}.
\end{barticle}
\endbibitem

\bibitem[\protect\citeauthoryear{{Kobanov}, {Chelpanov}, and
  {Kolobov}}{2013}]{Kobanov13}
\begin{barticle}
\bauthor{\bsnm{{Kobanov}}, \binits{N.I.}},
\bauthor{\bsnm{{Chelpanov}}, \binits{A.A.}},
\bauthor{\bsnm{{Kolobov}}, \binits{D.Y.}}:
\byear{2013},
\batitle{{Oscillations above sunspots from the temperature minimum to the
  corona}}.
\bjtitle{\aap}
\bvolume{554},
\bfpage{A146}.
\doiurl{10.1051/0004-6361/201220548}.
\adsurl{2013A\%26A...554A.146K}.
\end{barticle}
\endbibitem

\bibitem[\protect\citeauthoryear{{Kobanov} \textit{et~al.}}{2013}]{Kobanov13SP}
\begin{barticle}
\bauthor{\bsnm{{Kobanov}}, \binits{N.}},
\bauthor{\bsnm{{Kolobov}}, \binits{D.}},
\bauthor{\bsnm{{Kustov}}, \binits{A.}},
\bauthor{\bsnm{{Chupin}}, \binits{S.}},
\bauthor{\bsnm{{Chelpanov}}, \binits{A.}}:
\byear{2013},
\batitle{{Direct Measurement Results of the Time Lag of LOS-Velocity
  Oscillations Between Two Heights in Solar Faculae and Sunspots}}.
\bjtitle{\solphys}
\bvolume{284},
\bfpage{379}\,--\,\blpage{396}.
\doiurl{10.1007/s11207-013-0247-2}.
\adsurl{2013SoPh..284..379K}.
\end{barticle}
\endbibitem

\bibitem[\protect\citeauthoryear{{Kolotkov}, {Anfinogentov}, and
  {Nakariakov}}{2016}]{Kolotkov16}
\begin{barticle}
\bauthor{\bsnm{{Kolotkov}}, \binits{D.Y.}},
\bauthor{\bsnm{{Anfinogentov}}, \binits{S.A.}},
\bauthor{\bsnm{{Nakariakov}}, \binits{V.M.}}:
\byear{2016},
\batitle{{Empirical mode decomposition analysis of random processes in the
  solar atmosphere}}.
\bjtitle{\aap}
\bvolume{592},
\bfpage{A153}.
\doiurl{10.1051/0004-6361/201628306}.
\adsurl{2016A\%26A...592A.153K}.
\end{barticle}
\endbibitem

\bibitem[\protect\citeauthoryear{{Krishna Prasad}, {Jess}, and
  {Khomenko}}{2015}]{Prasad15}
\begin{barticle}
\bauthor{\bsnm{{Krishna Prasad}}, \binits{S.}},
\bauthor{\bsnm{{Jess}}, \binits{D.B.}},
\bauthor{\bsnm{{Khomenko}}, \binits{E.}}:
\byear{2015},
\batitle{{On the Source of Propagating Slow Magnetoacoustic Waves in
  Sunspots}}.
\bjtitle{\apjl}
\bvolume{812},
\bfpage{L15}.
\doiurl{10.1088/2041-8205/812/1/L15}.
\adsurl{2015ApJ...812L..15K}.
\end{barticle}
\endbibitem

\bibitem[\protect\citeauthoryear{{Kupriyanova}
  \textit{et~al.}}{2010}]{Kupriyanova10}
\begin{barticle}
\bauthor{\bsnm{{Kupriyanova}}, \binits{E.G.}},
\bauthor{\bsnm{{Melnikov}}, \binits{V.F.}},
\bauthor{\bsnm{{Nakariakov}}, \binits{V.M.}},
\bauthor{\bsnm{{Shibasaki}}, \binits{K.}}:
\byear{2010},
\batitle{{Types of Microwave Quasi-Periodic Pulsations in Single Flaring
  Loops}}.
\bjtitle{\solphys}
\bvolume{267},
\bfpage{329}\,--\,\blpage{342}.
\doiurl{10.1007/s11207-010-9642-0}.
\adsurl{2010SoPh..267..329K}.
\end{barticle}
\endbibitem

\bibitem[\protect\citeauthoryear{{Kupriyanova}
  \textit{et~al.}}{2019}]{Kupriyanova2019}
\begin{barticle}
\bauthor{\bsnm{{Kupriyanova}}, \binits{E.G.}},
\bauthor{\bsnm{{Kashapova}}, \binits{L.K.}},
\bauthor{\bsnm{{Van Doorsselaere}}, \binits{T.}},
\bauthor{\bsnm{{Chowdhury}}, \binits{P.}},
\bauthor{\bsnm{{Srivastava}}, \binits{A.K.}},
\bauthor{\bsnm{{Moon}}, \binits{Y.-J.}}:
\byear{2019},
\batitle{{Quasi-periodic pulsations in a solar flare with an unusual phase
  shift}}.
\bjtitle{\mnras}
\bvolume{483},
\bfpage{5499}\,--\,\blpage{5507}.
\doiurl{10.1093/mnras/sty3480}.
\adsurl{2019MNRAS.483.5499K}.
\end{barticle}
\endbibitem

\bibitem[\protect\citeauthoryear{{Kuznetsov}
  \textit{et~al.}}{2016}]{Kuznetsov16}
\begin{barticle}
\bauthor{\bsnm{{Kuznetsov}}, \binits{S.A.}},
\bauthor{\bsnm{{Zimovets}}, \binits{I.V.}},
\bauthor{\bsnm{{Morgachev}}, \binits{A.S.}},
\bauthor{\bsnm{{Struminsky}}, \binits{A.B.}}:
\byear{2016},
\batitle{{Spatio-temporal Dynamics of Sources of Hard X-Ray Pulsations in Solar
  Flares}}.
\bjtitle{\solphys}
\bvolume{291},
\bfpage{3385}\,--\,\blpage{3426}.
\doiurl{10.1007/s11207-016-0981-3}.
\adsurl{2016SoPh..291.3385K}.
\end{barticle}
\endbibitem

\bibitem[\protect\citeauthoryear{{Leenaarts}, {Carlsson}, and {Rouppe van der
  Voort}}{2012}]{Leenaarts12}
\begin{barticle}
\bauthor{\bsnm{{Leenaarts}}, \binits{J.}},
\bauthor{\bsnm{{Carlsson}}, \binits{M.}},
\bauthor{\bsnm{{Rouppe van der Voort}}, \binits{L.}}:
\byear{2012},
\batitle{{The Formation of the H{$\alpha$} Line in the Solar Chromosphere}}.
\bjtitle{\apj}
\bvolume{749},
\bfpage{136}.
\doiurl{10.1088/0004-637X/749/2/136}.
\adsurl{2012ApJ...749..136L}.
\end{barticle}
\endbibitem

\bibitem[\protect\citeauthoryear{{Lemen} \textit{et~al.}}{2012}]{sdoaia}
\begin{barticle}
\bauthor{\bsnm{{Lemen}}, \binits{J.R.}},
\bauthor{\bsnm{{Title}}, \binits{A.M.}},
\bauthor{\bsnm{{Akin}}, \binits{D.J.}},
\bauthor{\bsnm{{Boerner}}, \binits{P.F.}},
\bauthor{\bsnm{{Chou}}, \binits{C.}},
\bauthor{\bsnm{{Drake}}, \binits{J.F.}},
\bauthor{\bsnm{{Duncan}}, \binits{D.W.}},
\bauthor{\bsnm{{Edwards}}, \binits{C.G.}},
\bauthor{\bsnm{{Friedlaender}}, \binits{F.M.}},
\bauthor{\bsnm{{Heyman}}, \binits{G.F.}},
\bauthor{\bsnm{{Hurlburt}}, \binits{N.E.}},
\bauthor{\bsnm{{Katz}}, \binits{N.L.}},
\bauthor{\bsnm{{Kushner}}, \binits{G.D.}},
\bauthor{\bsnm{{Levay}}, \binits{M.}},
\bauthor{\bsnm{{Lindgren}}, \binits{R.W.}},
\bauthor{\bsnm{{Mathur}}, \binits{D.P.}},
\bauthor{\bsnm{{McFeaters}}, \binits{E.L.}},
\bauthor{\bsnm{{Mitchell}}, \binits{S.}},
\bauthor{\bsnm{{Rehse}}, \binits{R.A.}},
\bauthor{\bsnm{{Schrijver}}, \binits{C.J.}},
\bauthor{\bsnm{{Springer}}, \binits{L.A.}},
\bauthor{\bsnm{{Stern}}, \binits{R.A.}},
\bauthor{\bsnm{{Tarbell}}, \binits{T.D.}},
\bauthor{\bsnm{{Wuelser}}, \binits{J.-P.}},
\bauthor{\bsnm{{Wolfson}}, \binits{C.J.}},
\bauthor{\bsnm{{Yanari}}, \binits{C.}},
\bauthor{\bsnm{{Bookbinder}}, \binits{J.A.}},
\bauthor{\bsnm{{Cheimets}}, \binits{P.N.}},
\bauthor{\bsnm{{Caldwell}}, \binits{D.}},
\bauthor{\bsnm{{Deluca}}, \binits{E.E.}},
\bauthor{\bsnm{{Gates}}, \binits{R.}},
\bauthor{\bsnm{{Golub}}, \binits{L.}},
\bauthor{\bsnm{{Park}}, \binits{S.}},
\bauthor{\bsnm{{Podgorski}}, \binits{W.A.}},
\bauthor{\bsnm{{Bush}}, \binits{R.I.}},
\bauthor{\bsnm{{Scherrer}}, \binits{P.H.}},
\bauthor{\bsnm{{Gummin}}, \binits{M.A.}},
\bauthor{\bsnm{{Smith}}, \binits{P.}},
\bauthor{\bsnm{{Auker}}, \binits{G.}},
\bauthor{\bsnm{{Jerram}}, \binits{P.}},
\bauthor{\bsnm{{Pool}}, \binits{P.}},
\bauthor{\bsnm{{Soufli}}, \binits{R.}},
\bauthor{\bsnm{{Windt}}, \binits{D.L.}},
\bauthor{\bsnm{{Beardsley}}, \binits{S.}},
\bauthor{\bsnm{{Clapp}}, \binits{M.}},
\bauthor{\bsnm{{Lang}}, \binits{J.}},
\bauthor{\bsnm{{Waltham}}, \binits{N.}}:
\byear{2012},
\batitle{{The Atmospheric Imaging Assembly (AIA) on the Solar Dynamics
  Observatory (SDO)}}.
\bjtitle{\solphys}
\bvolume{275},
\bfpage{17}\,--\,\blpage{40}.
\doiurl{10.1007/s11207-011-9776-8}.
\adsurl{2012SoPh..275...17L}.
\end{barticle}
\endbibitem

\bibitem[\protect\citeauthoryear{{Lin} \textit{et~al.}}{2002}]{rhessi1}
\begin{barticle}
\bauthor{\bsnm{{Lin}}, \binits{R.P.}},
\bauthor{\bsnm{{Dennis}}, \binits{B.R.}},
\bauthor{\bsnm{{Hurford}}, \binits{G.J.}},
\bauthor{\bsnm{{Smith}}, \binits{D.M.}},
\bauthor{\bsnm{{Zehnder}}, \binits{A.}},
\bauthor{\bsnm{{Harvey}}, \binits{P.R.}},
\bauthor{\bsnm{{Curtis}}, \binits{D.W.}},
\bauthor{\bsnm{{Pankow}}, \binits{D.}},
\bauthor{\bsnm{{Turin}}, \binits{P.}},
\bauthor{\bsnm{{Bester}}, \binits{M.}},
\bauthor{\bsnm{{Csillaghy}}, \binits{A.}},
\bauthor{\bsnm{{Lewis}}, \binits{M.}},
\bauthor{\bsnm{{Madden}}, \binits{N.}},
\bauthor{\bsnm{{van Beek}}, \binits{H.F.}},
\bauthor{\bsnm{{Appleby}}, \binits{M.}},
\bauthor{\bsnm{{Raudorf}}, \binits{T.}},
\bauthor{\bsnm{{McTiernan}}, \binits{J.}},
\bauthor{\bsnm{{Ramaty}}, \binits{R.}},
\bauthor{\bsnm{{Schmahl}}, \binits{E.}},
\bauthor{\bsnm{{Schwartz}}, \binits{R.}},
\bauthor{\bsnm{{Krucker}}, \binits{S.}},
\bauthor{\bsnm{{Abiad}}, \binits{R.}},
\bauthor{\bsnm{{Quinn}}, \binits{T.}},
\bauthor{\bsnm{{Berg}}, \binits{P.}},
\bauthor{\bsnm{{Hashii}}, \binits{M.}},
\bauthor{\bsnm{{Sterling}}, \binits{R.}},
\bauthor{\bsnm{{Jackson}}, \binits{R.}},
\bauthor{\bsnm{{Pratt}}, \binits{R.}},
\bauthor{\bsnm{{Campbell}}, \binits{R.D.}},
\bauthor{\bsnm{{Malone}}, \binits{D.}},
\bauthor{\bsnm{{Landis}}, \binits{D.}},
\bauthor{\bsnm{{Barrington-Leigh}}, \binits{C.P.}},
\bauthor{\bsnm{{Slassi-Sennou}}, \binits{S.}},
\bauthor{\bsnm{{Cork}}, \binits{C.}},
\bauthor{\bsnm{{Clark}}, \binits{D.}},
\bauthor{\bsnm{{Amato}}, \binits{D.}},
\bauthor{\bsnm{{Orwig}}, \binits{L.}},
\bauthor{\bsnm{{Boyle}}, \binits{R.}},
\bauthor{\bsnm{{Banks}}, \binits{I.S.}},
\bauthor{\bsnm{{Shirey}}, \binits{K.}},
\bauthor{\bsnm{{Tolbert}}, \binits{A.K.}},
\bauthor{\bsnm{{Zarro}}, \binits{D.}},
\bauthor{\bsnm{{Snow}}, \binits{F.}},
\bauthor{\bsnm{{Thomsen}}, \binits{K.}},
\bauthor{\bsnm{{Henneck}}, \binits{R.}},
\bauthor{\bsnm{{McHedlishvili}}, \binits{A.}},
\bauthor{\bsnm{{Ming}}, \binits{P.}},
\bauthor{\bsnm{{Fivian}}, \binits{M.}},
\bauthor{\bsnm{{Jordan}}, \binits{J.}},
\bauthor{\bsnm{{Wanner}}, \binits{R.}},
\bauthor{\bsnm{{Crubb}}, \binits{J.}},
\bauthor{\bsnm{{Preble}}, \binits{J.}},
\bauthor{\bsnm{{Matranga}}, \binits{M.}},
\bauthor{\bsnm{{Benz}}, \binits{A.}},
\bauthor{\bsnm{{Hudson}}, \binits{H.}},
\bauthor{\bsnm{{Canfield}}, \binits{R.C.}},
\bauthor{\bsnm{{Holman}}, \binits{G.D.}},
\bauthor{\bsnm{{Crannell}}, \binits{C.}},
\bauthor{\bsnm{{Kosugi}}, \binits{T.}},
\bauthor{\bsnm{{Emslie}}, \binits{A.G.}},
\bauthor{\bsnm{{Vilmer}}, \binits{N.}},
\bauthor{\bsnm{{Brown}}, \binits{J.C.}},
\bauthor{\bsnm{{Johns-Krull}}, \binits{C.}},
\bauthor{\bsnm{{Aschwanden}}, \binits{M.}},
\bauthor{\bsnm{{Metcalf}}, \binits{T.}},
\bauthor{\bsnm{{Conway}}, \binits{A.}}:
\byear{2002},
\batitle{{The Reuven Ramaty High-Energy Solar Spectroscopic Imager (RHESSI)}}.
\bjtitle{\solphys}
\bvolume{210},
\bfpage{3}\,--\,\blpage{32}.
\doiurl{10.1023/A:1022428818870}.
\adsurl{2002SoPh..210....3L}.
\end{barticle}
\endbibitem

\bibitem[\protect\citeauthoryear{{Liu} \textit{et~al.}}{2011}]{Liu11}
\begin{barticle}
\bauthor{\bsnm{{Liu}}, \binits{W.}},
\bauthor{\bsnm{{Title}}, \binits{A.M.}},
\bauthor{\bsnm{{Zhao}}, \binits{J.}},
\bauthor{\bsnm{{Ofman}}, \binits{L.}},
\bauthor{\bsnm{{Schrijver}}, \binits{C.J.}},
\bauthor{\bsnm{{Aschwanden}}, \binits{M.J.}},
\bauthor{\bsnm{{De Pontieu}}, \binits{B.}},
\bauthor{\bsnm{{Tarbell}}, \binits{T.D.}}:
\byear{2011},
\batitle{{Direct Imaging of Quasi-periodic Fast Propagating Waves of \~{}2000
  km s$^{-1}$ in the Low Solar Corona by the Solar Dynamics Observatory
  Atmospheric Imaging Assembly}}.
\bjtitle{\apjl}
\bvolume{736},
\bfpage{L13}.
\doiurl{10.1088/2041-8205/736/1/L13}.
\adsurl{2011ApJ...736L..13L}.
\end{barticle}
\endbibitem

\bibitem[\protect\citeauthoryear{{McLaughlin}
  \textit{et~al.}}{2018}]{McLaughlin}
\begin{barticle}
\bauthor{\bsnm{{McLaughlin}}, \binits{J.A.}},
\bauthor{\bsnm{{Nakariakov}}, \binits{V.M.}},
\bauthor{\bsnm{{Dominique}}, \binits{M.}},
\bauthor{\bsnm{{Jel{\'{\i}}nek}}, \binits{P.}},
\bauthor{\bsnm{{Takasao}}, \binits{S.}}:
\byear{2018},
\batitle{{Modelling Quasi-Periodic Pulsations in Solar and Stellar Flares}}.
\bjtitle{\ssr}
\bvolume{214},
\bfpage{45}.
\doiurl{10.1007/s11214-018-0478-5}.
\adsurl{2018SSRv..214...45M}.
\end{barticle}
\endbibitem

\bibitem[\protect\citeauthoryear{{Milligan} \textit{et~al.}}{2017}]{Milligan17}
\begin{barticle}
\bauthor{\bsnm{{Milligan}}, \binits{R.O.}},
\bauthor{\bsnm{{Fleck}}, \binits{B.}},
\bauthor{\bsnm{{Ireland}}, \binits{J.}},
\bauthor{\bsnm{{Fletcher}}, \binits{L.}},
\bauthor{\bsnm{{Dennis}}, \binits{B.R.}}:
\byear{2017},
\batitle{{Detection of Three-minute Oscillations in Full-disk Ly{$\alpha$}
  Emission during a Solar Flare}}.
\bjtitle{\apjl}
\bvolume{848},
\bfpage{L8}.
\doiurl{10.3847/2041-8213/aa8f3a}.
\adsurl{2017ApJ...848L...8M}.
\end{barticle}
\endbibitem

\bibitem[\protect\citeauthoryear{{Nakariakov} and
  {Melnikov}}{2009}]{NakariakovMelnikov09}
\begin{barticle}
\bauthor{\bsnm{{Nakariakov}}, \binits{V.M.}},
\bauthor{\bsnm{{Melnikov}}, \binits{V.F.}}:
\byear{2009},
\batitle{{Quasi-Periodic Pulsations in Solar Flares}}.
\bjtitle{\ssr}
\bvolume{149},
\bfpage{119}\,--\,\blpage{151}.
\doiurl{10.1007/s11214-009-9536-3}.
\adsurl{2009SSRv..149..119N}.
\end{barticle}
\endbibitem

\bibitem[\protect\citeauthoryear{{Nakariakov} and
  {Zimovets}}{2011}]{NakariakovZimovets11}
\begin{barticle}
\bauthor{\bsnm{{Nakariakov}}, \binits{V.M.}},
\bauthor{\bsnm{{Zimovets}}, \binits{I.V.}}:
\byear{2011},
\batitle{{Slow Magnetoacoustic Waves in Two-ribbon Flares}}.
\bjtitle{\apjl}
\bvolume{730},
\bfpage{L27}.
\doiurl{10.1088/2041-8205/730/2/L27}.
\adsurl{2011ApJ...730L..27N}.
\end{barticle}
\endbibitem

\bibitem[\protect\citeauthoryear{{Nakariakov}
  \textit{et~al.}}{2016}]{Nakariakov2016SSR}
\begin{barticle}
\bauthor{\bsnm{{Nakariakov}}, \binits{V.M.}},
\bauthor{\bsnm{{Pilipenko}}, \binits{V.}},
\bauthor{\bsnm{{Heilig}}, \binits{B.}},
\bauthor{\bsnm{{Jel{\'{\i}}nek}}, \binits{P.}},
\bauthor{\bsnm{{Karlick{\'y}}}, \binits{M.}},
\bauthor{\bsnm{{Klimushkin}}, \binits{D.Y.}},
\bauthor{\bsnm{{Kolotkov}}, \binits{D.Y.}},
\bauthor{\bsnm{{Lee}}, \binits{D.-H.}},
\bauthor{\bsnm{{Nistic{\`o}}}, \binits{G.}},
\bauthor{\bsnm{{Van Doorsselaere}}, \binits{T.}},
\bauthor{\bsnm{{Verth}}, \binits{G.}},
\bauthor{\bsnm{{Zimovets}}, \binits{I.V.}}:
\byear{2016},
\batitle{{Magnetohydrodynamic Oscillations in the Solar Corona and Earth's
  Magnetosphere: Towards Consolidated Understanding}}.
\bjtitle{\ssr}
\bvolume{200},
\bfpage{75}\,--\,\blpage{203}.
\doiurl{10.1007/s11214-015-0233-0}.
\adsurl{2016SSRv..200...75N}.
\end{barticle}
\endbibitem

\bibitem[\protect\citeauthoryear{{Ning}}{2014}]{Ning}
\begin{barticle}
\bauthor{\bsnm{{Ning}}, \binits{Z.}}:
\byear{2014},
\batitle{{Imaging Observations of X-Ray Quasi-periodic Oscillations at 3 - 6
  keV in the 26 December 2002 Solar Flare}}.
\bjtitle{\solphys}
\bvolume{289},
\bfpage{1239}\,--\,\blpage{1256}.
\doiurl{10.1007/s11207-013-0405-6}.
\adsurl{2014SoPh..289.1239N}.
\end{barticle}
\endbibitem

\bibitem[\protect\citeauthoryear{{Nistic{\`o}}, {Pascoe}, and
  {Nakariakov}}{2014}]{Nistico2014}
\begin{barticle}
\bauthor{\bsnm{{Nistic{\`o}}}, \binits{G.}},
\bauthor{\bsnm{{Pascoe}}, \binits{D.J.}},
\bauthor{\bsnm{{Nakariakov}}, \binits{V.M.}}:
\byear{2014},
\batitle{{Observation of a high-quality quasi-periodic rapidly propagating wave
  train using SDO/AIA}}.
\bjtitle{\aap}
\bvolume{569},
\bfpage{A12}.
\doiurl{10.1051/0004-6361/201423763}.
\adsurl{2014A\%26A...569A..12N}.
\end{barticle}
\endbibitem

\bibitem[\protect\citeauthoryear{{Pesnell}, {Thompson}, and
  {Chamberlin}}{2012}]{Pesnell12}
\begin{barticle}
\bauthor{\bsnm{{Pesnell}}, \binits{W.D.}},
\bauthor{\bsnm{{Thompson}}, \binits{B.J.}},
\bauthor{\bsnm{{Chamberlin}}, \binits{P.C.}}:
\byear{2012},
\batitle{{The Solar Dynamics Observatory (SDO)}}.
\bjtitle{\solphys}
\bvolume{275},
\bfpage{3}\,--\,\blpage{15}.
\doiurl{10.1007/s11207-011-9841-3}.
\adsurl{2012SoPh..275....3P}.
\end{barticle}
\endbibitem

\bibitem[\protect\citeauthoryear{{Pugh} \textit{et~al.}}{2017}]{Pugh17}
\begin{barticle}
\bauthor{\bsnm{{Pugh}}, \binits{C.E.}},
\bauthor{\bsnm{{Nakariakov}}, \binits{V.M.}},
\bauthor{\bsnm{{Broomhall}}, \binits{A.-M.}},
\bauthor{\bsnm{{Bogomolov}}, \binits{A.V.}},
\bauthor{\bsnm{{Myagkova}}, \binits{I.N.}}:
\byear{2017},
\batitle{{Properties of quasi-periodic pulsations in solar flares from a single
  active region}}.
\bjtitle{\aap}
\bvolume{608},
\bfpage{A101}.
\doiurl{10.1051/0004-6361/201731636}.
\adsurl{2017A\%26A...608A.101P}.
\end{barticle}
\endbibitem

\bibitem[\protect\citeauthoryear{{Rayrole}}{1967}]{Rayrole}
\begin{barticle}
\bauthor{\bsnm{{Rayrole}}, \binits{J.}}:
\byear{1967},
\batitle{{Contribution {\`a} l'{\'e}tude de la structure du champ
  magn{\'e}tique dans les taches solaires}}.
\bjtitle{Annales d'Astrophysique}
\bvolume{30},
\bfpage{257}.
\adsurl{1967AnAp...30..257R}.
\end{barticle}
\endbibitem

\bibitem[\protect\citeauthoryear{{Sych} \textit{et~al.}}{2009}]{Sych2009}
\begin{barticle}
\bauthor{\bsnm{{Sych}}, \binits{R.}},
\bauthor{\bsnm{{Nakariakov}}, \binits{V.M.}},
\bauthor{\bsnm{{Karlicky}}, \binits{M.}},
\bauthor{\bsnm{{Anfinogentov}}, \binits{S.}}:
\byear{2009},
\batitle{{Relationship between wave processes in sunspots and quasi-periodic
  pulsations in active region flares}}.
\bjtitle{\aap}
\bvolume{505},
\bfpage{791}\,--\,\blpage{799}.
\doiurl{10.1051/0004-6361/200912132}.
\adsurl{2009A\%26A...505..791S}.
\end{barticle}
\endbibitem

\bibitem[\protect\citeauthoryear{{Terekhov}
  \textit{et~al.}}{2002}]{Terekhov2002}
\begin{barticle}
\bauthor{\bsnm{{Terekhov}}, \binits{O.V.}},
\bauthor{\bsnm{{Shevchenko}}, \binits{A.V.}},
\bauthor{\bsnm{{Kuz'min}}, \binits{A.G.}},
\bauthor{\bsnm{{Sazonov}}, \binits{S.Y.}},
\bauthor{\bsnm{{Sunyaev}}, \binits{R.A.}},
\bauthor{\bsnm{{Lund}}, \binits{N.}}:
\byear{2002},
\batitle{{Observation of Quasi-Periodic Pulsations in the Solar Flare SF
  900610}}.
\bjtitle{Astronomy Letters}
\bvolume{28},
\bfpage{397}\,--\,\blpage{400}.
\doiurl{10.1134/1.1484140}.
\adsurl{2002AstL...28..397T}.
\end{barticle}
\endbibitem

\bibitem[\protect\citeauthoryear{{Yuan} \textit{et~al.}}{2013}]{YuanShen2013}
\begin{barticle}
\bauthor{\bsnm{{Yuan}}, \binits{D.}},
\bauthor{\bsnm{{Shen}}, \binits{Y.}},
\bauthor{\bsnm{{Liu}}, \binits{Y.}},
\bauthor{\bsnm{{Nakariakov}}, \binits{V.M.}},
\bauthor{\bsnm{{Tan}}, \binits{B.}},
\bauthor{\bsnm{{Huang}}, \binits{J.}}:
\byear{2013},
\batitle{{Distinct propagating fast wave trains associated with flaring energy
  releases}}.
\bjtitle{\aap}
\bvolume{554},
\bfpage{A144}.
\doiurl{10.1051/0004-6361/201321435}.
\adsurl{2013A\%26A...554A.144Y}.
\end{barticle}
\endbibitem

\bibitem[\protect\citeauthoryear{{Zhang}
  \textit{et~al.}}{2015}]{ZhangZhang2015}
\begin{barticle}
\bauthor{\bsnm{{Zhang}}, \binits{Y.}},
\bauthor{\bsnm{{Zhang}}, \binits{J.}},
\bauthor{\bsnm{{Wang}}, \binits{J.}},
\bauthor{\bsnm{{Nakariakov}}, \binits{V.M.}}:
\byear{2015},
\batitle{{Coexisting fast and slow propagating waves of the extreme-UV
  intensity in solar coronal plasma structures}}.
\bjtitle{\aap}
\bvolume{581},
\bfpage{A78}.
\doiurl{10.1051/0004-6361/201525621}.
\adsurl{2015A\%26A...581A..78Z}.
\end{barticle}
\endbibitem

\bibitem[\protect\citeauthoryear{{Zhao} \textit{et~al.}}{2016}]{Zhao16}
\begin{barticle}
\bauthor{\bsnm{{Zhao}}, \binits{J.}},
\bauthor{\bsnm{{Felipe}}, \binits{T.}},
\bauthor{\bsnm{{Chen}}, \binits{R.}},
\bauthor{\bsnm{{Khomenko}}, \binits{E.}}:
\byear{2016},
\batitle{{Tracing p-mode Waves from the Photosphere to the Corona in Active
  Regions}}.
\bjtitle{\apjl}
\bvolume{830},
\bfpage{L17}.
\doiurl{10.3847/2041-8205/830/1/L17}.
\adsurl{2016ApJ...830L..17Z}.
\end{barticle}
\endbibitem

\bibitem[\protect\citeauthoryear{{Zimovets} and
  {Struminsky}}{2009}]{Zimovets09}
\begin{barticle}
\bauthor{\bsnm{{Zimovets}}, \binits{I.V.}},
\bauthor{\bsnm{{Struminsky}}, \binits{A.B.}}:
\byear{2009},
\batitle{{Imaging Observations of Quasi-Periodic Pulsatory Nonthermal Emission
  in Two-Ribbon Solar Flares}}.
\bjtitle{\solphys}
\bvolume{258},
\bfpage{69}\,--\,\blpage{88}.
\doiurl{10.1007/s11207-009-9394-x}.
\adsurl{2009SoPh..258...69Z}.
\end{barticle}
\endbibitem

\end{thebibliography}

\end{document}